\documentstyle[12pt,frascatiphys1,amsmath,amssymb]{article}

\def\dd{{\mathrm d}}

\begin{document}
\title{THE LIGHTEST STRANGE SCALAR MESON\thanks{Talk presented by Stuart Cherry}}
\author{Stuart Cherry \\
{\em University Of Durham, Durham, DH1 3LE, U.K.} \\
M.R.~Pennington \\
{\em University Of Durham, Durham, DH1 3LE, U.K.}}

\maketitle
\baselineskip=14.5pt
\begin{abstract}
I present the results of a recent calculation to determine the number of strange scalar resonances below 1.8~GeV based on the analytic properties of the experimental $\pi K$ scattering amplitude.
Only one resonance was found in the data, and this is readily identifiable as the $K_0^*(1430)$.
We found no evidence to support the $\kappa(900)$.
\end{abstract}
\baselineskip=17pt

\section{Introduction}
The scalar mesons are one of the most controversial subjects in hadron physics.
In the last edition of the PDG\cite{Caso}, there were four scalar isoscalars listed below 1.5~GeV (the $f_0(400-1200)$, $f_0(980)$,  $f_0(1370)$ and $f_0(1500)$, with a possible fifth at 1.7~GeV, the $f_J(1710)$.
There were also two isovectors listed, the $a_0(980)$ and the $a_0(1450)$.
This is obviously too many for a standard $q{\overline q}$ nonet.
However, many QCD-motivated models predict the the existence of non-$q{\overline q}$ mesons, such as $qq{\overline {qq}}$ states\cite{Jaffe1}, $K{\overline K}$ molecules\cite{Weinstein} and glueballs\cite{Klempt}.
It is precisely in the scalar-isoscalar that these unconventional mesons are most like to be found.
This excess of isoscalars and isovectors has led to the suggestion that there are in fact two scalar nonets\cite{Black2}: 
the conventional one centred around 1.4~GeV and an unconventional, possibly $qq{\overline {qq}}$, one centred around 1~GeV.
However, the PDG lists only one pair of strange scalar mesons in this region, the $K^*_0(1430)$ and so some authors have postulated a light strange meson, known as the $\kappa(900)$ 
Evidence for\cite{Ishida} and against\cite{Anisovich} the $\kappa(900)$ has been claimed within various models.
It should be kept in mind that the existence of a resonance is not defined by the ability to fit a particular formula to the experimental data.
A resonance is entirely defined by the presence of a pole of the $S-$matrix on the nearby unphysical sheet.

I would like to present to you the results of a model-independent calculation\cite{Cherry} to determine whether the $\kappa(900)$ is indeed present in the experimental data by determining the number and positions of the poles of the $S-$matrix.
The method, due to Nagov\'a {\it et al}\cite{Nagova}, does not require the artificial separation of data into background and resonance components and I will begin by briefly outlining it.

\section{Mapping}

\begin{figure}[!htb]
\begin{center}
\vspace{4cm}
\includegraphics{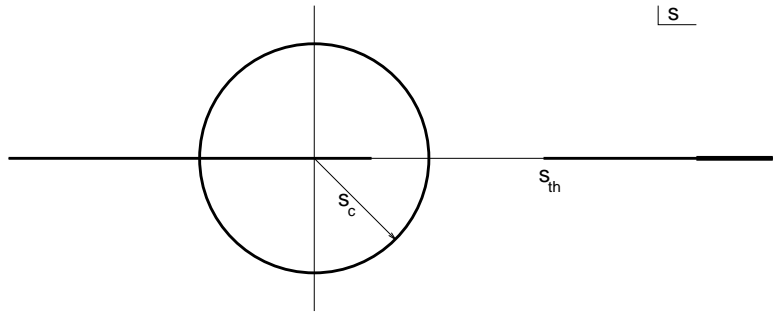}
\caption{\it The cut structure of the $\pi K$ partial wave amplitude, where \mbox{$s_{th}=(m_K+m_{\pi})^2$} and the radius of the circular cut is \mbox{$s_c=m_K^2-m_{\pi}^2$}.}
\label{PWA}
\end{center}
\vspace{-1cm}
\end{figure}

To determine the position of a pole in the complex plane from data along the real axis we must perform some form of analytic continuation.
In order to maximise the region of validity of our analytic continuation we begin by conformally mapping the unphysical sheet of the $\pi K$ partial wave amplitude (see Fig.~\ref{PWA}) into the unit disc.

\begin{figure}[!htb]
\begin{center}
\vspace{4cm}
\includegraphics{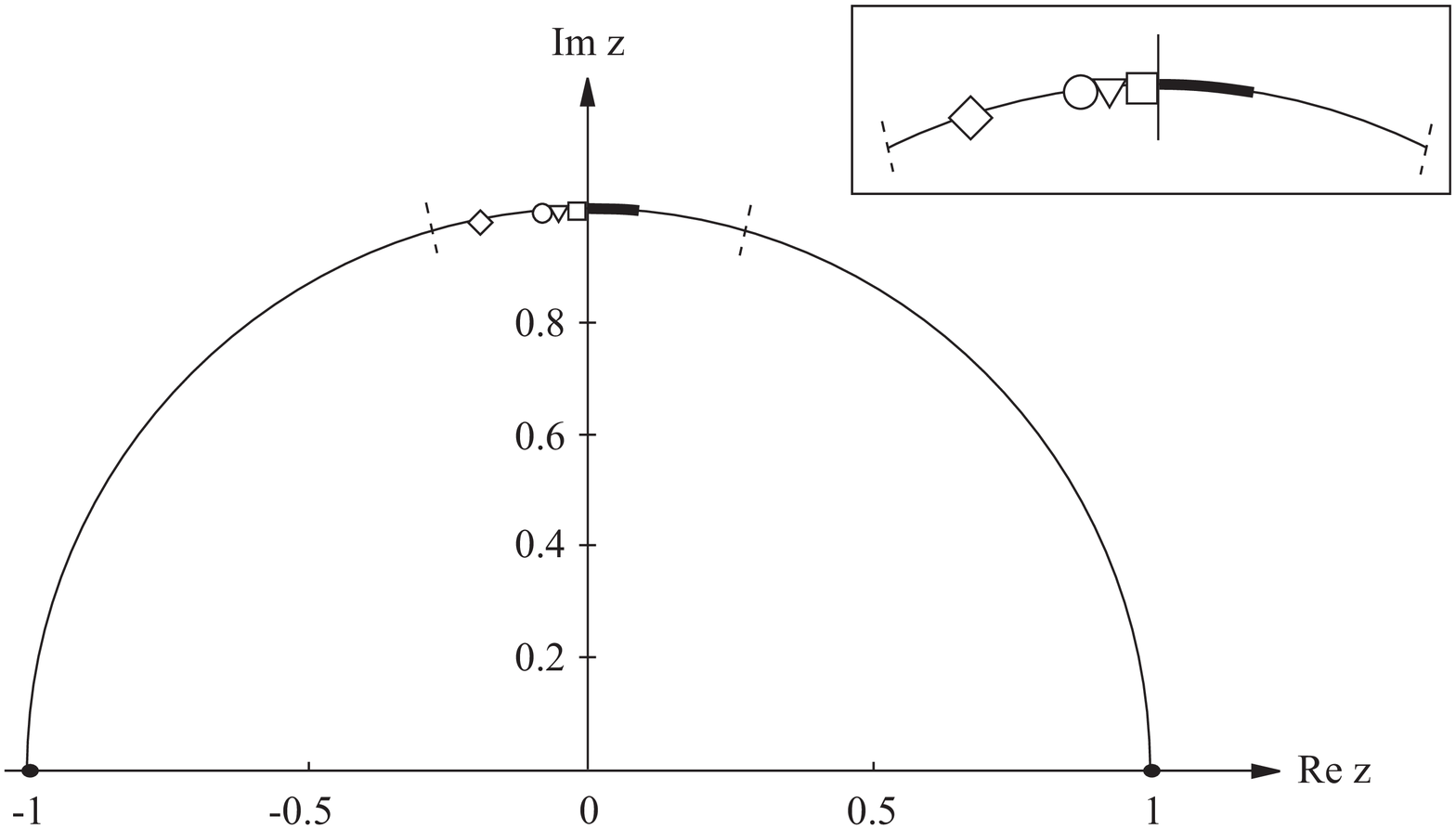}
\caption{\it The z-plane showing how points in s-plane map for a given $\beta$.
The thicker line shows the arc covered by the LASS data.
The symbols mark particular values of $s$: $ \blacklozenge~s~=~s_{th},\ \square~s~=~\infty,\  \triangledown \, s = -s_c,\   \circ \, s = i s_c,\  \lozenge \, s = \frac{1}{\sqrt{2}}(s_c + i s_c),\  \bullet \, s = s_c$.} 
\label{zcut}
\end{center}
\vspace{-1.5cm}
\end{figure}

The mapping is designed so that the cuts in the $s-$plane are mapped onto the circumference of the circle in the $z-$plane.
The mapping is accomplished in two steps.
First 
\begin{equation}
y(s) = \left( \frac{s-s_c}{s+s_c} \right) ^2\quad {\rm and \ then}\quad
z(s) = \frac{i \beta \sqrt{y(s)} - \sqrt{y(s)-y(s_{th})}}{i \beta \sqrt{y(s)} + \sqrt{y(s)-y(s_{th})}}\quad ,
\end{equation}
where $\beta$ is a real number chosen to minimise the distance the continuation must cover.
From Fig.~\ref{zcut} we notice that physical data could never cover the whole circle.
We also see that the proportion of the circle covered by each increment in $s$ falls very sharply as $s$ increases.
Hence we can neglect radial excitations from our analysis.

\section{Continuation}

Let $Y(z)$ and $\epsilon (z)$ be continuous functions describing the experimental data and errors around the entire circle.
If $F(z)$ is a square-integrable function around the circle, then we can test how well it describes the data through a $\chi^2$ defined by
\begin{equation}
\chi^2\, =\, \frac{1}{2\pi} \oint_C \bigg\vert \frac{F(z) - Y(z)}{\epsilon(z)}\bigg\vert^2 \, \vert \dd z \vert \quad .
\label{chi1}
\end{equation}

We introduce a weight function $g(z)$ which we define to be real analytic, non-zero throughout the disc and constrained by $\vert g(z) \vert = \epsilon (z)$ around the circle.
Expanding our data and trial functions as Laurent series about the origin, i.e. 
\begin{equation}
y(z) = \frac{Y(z)}{g(z)} = \sum_{k = -\infty}^{\infty} y_k\, z^k \qquad  {\rm and} \qquad
f(z) = \frac{F(z)}{g(z)} = \sum_{k = -\infty}^{\infty} a_k\, z^k\quad,
\label{ser}
\end{equation}
gives
\begin{equation}
\chi^2 = \sum_{k = 1}^{\infty} \left(a_{-k} - y_{-k}\right)^2 + \sum_{k = 0}^{\infty} \left(a_k - y_k\right)^2 
\label{chi2}
\end{equation}
The pole structure of our partial wave amplitude, $Y(z)$, can then be determined by finding the test function $F(z)$ which minimises the first summation in Eq.~(\ref{chi2}).
Since partial wave amplitudes are real analytic, the coefficients $a_k$ and $y_k$ will be real.

To test if the amplitude is free of resonances we set $a_{-k} \equiv 0 \ \forall \ k > 0$.
Then the quantity
\mbox{$\chi_0^2 \ = \sum_{k = 1}^{\infty} y_{-k}^2$}
will have a $\chi^2-$distribution.

To test if the amplitude contains one resonance we define 
\begin{equation}
\tilde{y} (z) = \frac{Y(z) B_{\lambda} (z)}{g(z)} = \sum_{k=-\infty}^{\infty} \tilde{y}_{k}\, z^k \quad {\rm where} \quad 
B_{\lambda} (z) = \frac{(z-\lambda)(z-\lambda^*)}{(1-z \lambda)(1-z \lambda^*)} \quad .
\end{equation}
$B_{\lambda}$ is the one resonance Blaschke pole killing factor.
Now the quantity
\mbox{$\chi_1^2 \ = \sum_{k = 1}^{\infty} \tilde{y}_{-k}^2$}
will have a $\chi^2-$distribution.

\section{Inputs}

The experimental inputs into our calculation are the $\pi K$ scattering magnitudes, $a(s)$, and phases, $\phi(s)$, as measured, for example, by the LASS experiment\cite{Aston}.
The amplitude is normalised such that
\begin{equation}
f^I(s) = \frac{a(s) e^{i \phi(s)}}{\rho(s)} \qquad , \quad \rho(s) = \frac{2 q(s)}{\sqrt{s}}
\end{equation}

\noindent where $q(s)$ is the centre of mass 3-momentum of the $\pi K$ system and the superscript indicates we are on the physical sheet.
In order to detect a resonance we must move onto the relevant unphysical sheet, this is done by swapping the sign of the phase.
After mapping we have discrete data points $f_i(z)$ and errors $\Delta_i$ around a portion of the circle.
Where the data is more densely packed the amplitude is more tightly controlled, so we weight the errors by the density of data points in that region, i.e. we define $\epsilon_i = \Delta_i \sqrt{\vert \delta z_i \vert/2 \pi}$.

Between threshold and the start of the data we use the LASS effective range fit to create a few guide points.
In the unphysical region we simply guess a few widely spaced guide points and assign large errors.
The large spacing and errors will hopefully deweight these points so that they do not unduly affect the results of the calculation.
With these three different inputs and using the Schwarz Reflection Principle we can cover the circle with data and then we interpolate to give the continuous functions that we need.

A suitable form for the weight function is
$g(z) = \exp{\sum_{n=1}^{M} c_n z^n}$
where the $c_n$ can be found from a Fourier cosine fit to $\log{\epsilon(z)}$.

We can now calculate the singular coefficients of our Laurent expansions in the usual way, i.e. 
\begin{equation}
y_k = \frac{1}{2 \pi} \oint_c \frac{Y(z) z^k}{g(z)} \vert \dd z \vert \qquad {\rm and} \qquad
\tilde{y}_k = \frac{1}{2 \pi} \oint_c \frac{Y(z) B_{\lambda}(z) z^k}{g(z)} \vert \dd z \vert
\end{equation}

We want to find the value of $\lambda$ that minimises $\chi_1^2$ and then compare the final value with $\chi_0^2$.
We can carry out a similar procedure to test for two resonances and by comparing the various $\chi^2$'s we can determine the number of resonances present in the channel.

\section{Results}

In order to assess its capabilities, we began by testing the method explained previously with a trial scattering amplitude.
This amplitude was created using Jost functions and had two resonances present, a $\kappa_1$ with mass 900~MeV and width 500~GeV and a $\kappa_2$ with mass 1400~MeV and width 300~MeV.
The results are shown in Table~\ref{JOST}.
We can see clearly that the $\chi^2$ falls significantly as the number of resonances is increased and that changing the size of the errors in the unphysical region has very little effect on either of the pole positions obtained.
Thus we can confidently claim that the amplitude contains two resonances and furthermore the parameters of these resonances have been quite accurately determined.

\begin{table}[!htb]
\begin{center}
\begin{tabular}{|c|c|c|c|}
\hline
Option & No. of & $\sqrt{s_{pole}}$ & $\chi^2$ \\
& resonances &  (MeV) & \\
\hline \hline
& 0  & -  & 4081~~\\
\cline{2-4}
1 & 1  & $1201 \pm 131 i$ & ~281~~\\
\cline{2-4}
& 2 & $1396 \pm 142 i$ $903 \pm 234 i$ & ~~0.5\\
\hline \hline
& 0 & - & 3759~~\\
\cline{2-4}
2 & 1 & $1173 \pm 111 i$ & ~417\\
\cline{2-4}
& 2  & $1392 \pm 140 i$ $900 \pm 233 i$& ~~1.0\\
\hline
\end{tabular}
\end{center}
\vspace{-5mm}
\caption{\it Pole positions and $\chi^2$'s for model data.  
The two options differ only in the size of the errors in the unphysical region.}
\label{JOST}
\end{table}

\begin{table}[!htb]
\begin{center}
\begin{tabular}{|c|c|c|c|}
\hline
Option & No. of   & $\sqrt{s_{pole}}$ & $\chi^2$ \\
& resonances  & (MeV) &\\
\hline \hline
  & 0  & - & 1373~~ \\
\cline{2-4}
1 & 1  & $1433 \pm 149 i$ & ~~4.7\\
\cline{2-4}
  & 2  & $1432 \pm 148 i$  $805 \pm 13 i$ & ~~1.8\\
\hline \hline
  & 0 & - & 1629~~\\
\cline{2-4}
2 & 1 & $1423 \pm 157 i$ & ~~9.0\\
\cline{2-4}
  & 2  & $1423 \pm 154 i$ $969 \pm 6 i$ & ~~8.8\\
\hline
\end{tabular}
\end{center}
\vspace{-5mm}
\caption{\it Pole positions and $\chi^2$'s for LASS $S-$wave data from 0.825~GeV to 2.51~GeV.
The two Options differ only in the size of the errors in the unphysical region.}
\label{LASS}
\end{table}

Turning now to experimental data, Table~\ref{LASS} shows the results obtained from the LASS total S-wave data.
We see that the fall in $\chi^2$ in going from one to two resonances is negligible compared to the fall between zero and one resonance.
Also the parameters for the second resonance are quite sensitive to the size of the errors in the unphysical region.
This suggests that the second resonance listed in Table~\ref{LASS} is merely an artifact of looking for two poles and is not really present in the data.

\section{Conclusions}

In this talk I have tried to outline a calculation to determine the number of resonances present in strange scalar channel.
This method is most sensitive to lighter resonances and is capable of identifying two resonances in a channel even when they are broad and overlapping.
The experimental $\pi K$ scattering data is found to contain only one resonance below 1.8~GeV  and this state is readily identifiable with the $K_0^*(1430)$.
There is no evidence for a $\kappa(900)$.

I would like to thank the organisers for the opportunity to attend such an interesting and enjoyable meeting.

\end{document}